\documentclass[
 aps, pra,
 amsmath,amssymb,
 11pt,
 final,
tightenlines,
 nofloats,
nofootinbib,
 superscriptaddress,
showkeys,
showkeywords,
 ]
{revtex4-2}

\usepackage[T2A]{fontenc}
\usepackage[utf8x]{inputenc}
\usepackage[russian,english]{babel}
\usepackage{graphicx}% Include figure files
\usepackage{dcolumn}% Align table columns on decimal point
\usepackage{bm}% bold math

\input{maik.rty}

\setcitestyle{authoryear,round,aysep={,},citesep={;}}
\def\squareforqed{\hbox{\rlap{$\sqcap$}$\sqcup$}}

\def\sq{\ifmmode\squareforqed\else{\unskip\nobreak\hfil
\penalty50\hskip1em\null\nobreak\hfil\squareforqed
\parfillskip=0pt\finalhyphendemerits=0\endgraf}\fi}

\def\utw{\smash{\rlap{\lower5pt\hbox{$\sim$}}}}

\def\udtw{\smash{\rlap{\lower6pt\hbox{$\approx$}}}}

\def\diameter{{\ifmmode\mathchoice
{\ooalign{\hfil\hbox{$\displaystyle/$}\hfil\crcr
{\hbox{$\displaystyle\mathchar"20D$}}}}
{\ooalign{\hfil\hbox{$\textstyle/$}\hfil\crcr
{\hbox{$\textstyle\mathchar"20D$}}}}
{\ooalign{\hfil\hbox{$\scriptstyle/$}\hfil\crcr
{\hbox{$\scriptstyle\mathchar"20D$}}}}
{\ooalign{\hfil\hbox{$\scriptscriptstyle/$}\hfil\crcr
{\hbox{$\scriptscriptstyle\mathchar"20D$}}}}
\else{\ooalign{\hfil/\hfil\crcr\mathhexbox20D}}%
\fi}}

\begin{document}

 
\title{Structure of the Stellar Neighborhood of CQ~Tau in the Presence of a Companion in an Elongated Orbit}

\author{\firstname{T.~V.}~\surname{Demidova}}
 \email{proxima1@list.ru}
 \affiliation{Crimean Astrophysical Observatory, Russian Academy of Sciences, Nauchny, 298409 Russia}
\begin{abstract}

There are indirect signs of the presence of a companion near the star CQ\,Tau. A period of 10\;years was found in the brightness changes. The image of the disk shows an extensive cavity with a size of about 25~AU, surrounded by a dust ring with a distribution maximum near 53~AU. A simulation of the interaction of dust and gas in the vicinity of the star with the CQ\,Tau parameters was carried out, assuming the existence of a companion in an orbit with a period of 10\;years. It was shown that an M-class star in a highly elongated orbit is capable of forming a region of reduced density around the center of mass of the system, with a size close to the observed one. However, the dust ring-like structure in this case is located noticeably closer to the star than observed. Evidence has been obtained that a massive planet in an outer, relatively binary orbit, could form a ring-shaped dust structure at a distance similar to that observed.
\end{abstract}

\maketitle

\section{INTRODUCTION}\label{sec:intro}

The star CQ\,Tau is one of the brightest and closest UX\,Ori type stars~\citep{2022MNRAS.515.6109H} and at the same time the most active member of this family. The photometric properties of the star and a brief history of its studies are described in detail in a recent paper~\citet{2023Ap.....66..235G}.  It shows that the brightness variations of CQ\,Tau have a period of 10\;years, which was previously suspected in 
\citet{2005Ap.....48..135S}. It indicates the existence of strong, periodic disturbances in the inner region of the circumstellar disk, caused by the motion of the companion.

An indirect indication of the existence of a companion is a large cavity in the circumstellar disk of the star, discovered by observations in the millimeter range using the ALMA\footnote{Atacama Large Millimeter Array} interferometer ~\citep{2019MNRAS.486.4638U,2017ApJ...845...44T,2021A&A...648A..19W}. Analysis of interferometric observations in the continuum (\mbox{$\lambda=1.3$\;mm}) showed that the cavity extends in the disk from 15 to 25\;AU from the center~\citep{2019MNRAS.486.4638U}. Since the disk radiation in the millimeter range is formed by large particles with a characteristic size of about 1\;mm, this cavity indicates a deficit of large particles. According to the same paper, the size of the gas cavity, determined by the lines of the CO molecule, is somewhat smaller. 

In~\citet{2019MNRAS.486.4638U} it was shown that a massive planet on a circular orbit with a semi-major axis \mbox{$a_p=20$} AU can clear a ring-shaped cavity whose outer boundary corresponds to that observed for CQ\,Tau. However, in the central part of the disk with a radius of about 15\;AU, dust and gas are preserved. Thus, a massive planet cannot form the cavity observed in the image of CQ\,Tau. In addition, the orbital period of the planet in such a model is \mbox{$P_p\approx67$\;years}, so it cannot explain the brightness variations of the star with a period of 10\;years.

In the paper, we consider a model in which the CQ\,Tau companion orbits the star with a period of \mbox{$P=10$\;yr}. In this case, the semi-major axis of the orbit will depend on the mass of the companion, but it is noticeably smaller than the size of the observed cavity (\mbox{$a\sim6$\;AU}). In~\citet{1994ApJ...421..651A}, it was shown that with increasing eccentricity, the size of the cavity that can be cleared by the binary system increases. In addition, periodic variations in the column density along the line of sight, caused by the dynamics of matter in the inner part of the disk relative to the orbit, are possible only in the case of an eccentric orbit of the binary system~\citep{2017AstL...43..106D}. Observations also show indirect evidence for the presence of an unseen close companion at a distance of \mbox{2--8\;AU} on an eccentric orbit~\citep{2022MNRAS.515.6109H}. Therefore, the simulation assumes that the companion orbit has a large eccentricity.

\section{MODEL}\label{sec:mod}

The paper analyzes a model of a system consisting of a star with a mass \mbox{$M_1=1.67\,M_\odot$}, radius \mbox{$R_1=2.2\,R_\odot$} and temperature \mbox{$T_1=6900$\;K} and a low-mass companion with a mass $M_2$ (varies), which are immersed in a gas-dust disk. An elongated orbit of the companion, the eccentricity of which also varies, with an orbital period of \mbox{$P=10$\;years} was assumed. At the initial moment of time, the star and companion were located in the apoaster on the $x$ axis, which was directed from the star to the companion. The origin of coordinates corresponds to the center of mass of the binary system.

In addition to the model described above, a model with a planet orbiting a binary system was also studied. In this case, a point mass $m_p$ was added to the system, which gravitationally affects the disk matter, but does not affect the motion of the binary star. The orbit of the planet is determined by the parameters: $a_p$~is semimajor axis, $e_p$~is eccentricity, $\omega_p$~is argument of pericenter. Its initial position is determined by the value of the true anomaly $f_p$. The orbit of the planet was assumed to be coplanar with the orbit of the binary system and the equatorial plane of the disk.

At the initial moment of time, the disk matter is distributed within the limits of \mbox{$r_{\rm in}=0.5$\;AU}, \mbox{$r_{\rm out}=100$\;AU} according to the standard density distribution law described in~\cite{1994A&A...286..149D}:
\begin{equation}
\rho(r,z,0)=\cfrac{\Sigma}{\sqrt{2\pi}H(r)}\exp\Bigg[-\cfrac{z^2}{2H^2(r)}\Bigg],
\label{Eq:rho}
\end{equation}
where $r$~ is the distance from the center of mass of the binary system in the plane of the disk (cylindrical radius), and $z$~ is the distance from the plane of the disk.

The surface density in the disk was determined in accordance with the law adopted in~\citet{2019MNRAS.486.4638U}:
\begin{equation}
\Sigma(r,0)=\Sigma(r_0,0)\bigg(\cfrac{r}{r_0}\bigg)^{-\gamma}\exp\Bigg[-\bigg(\cfrac{r}{r_0}\bigg)^{2-\gamma}\Bigg],
\label{Eq:sigmaGas}
\end{equation}
where \mbox{$\gamma=0.3$}, \mbox{$\Sigma(r_0,0)=2.5$~g\,cm$^{-2}$} for gas and \mbox{$\gamma=-0.7$}, \mbox{$\Sigma(r_0,0)=0.85$~g\,cm$^{-2}$} for dust of 1\;mm, \mbox{$r_0=56$\;AU}

The half-thickness of the disk for gas is determined by the formula
\begin{equation} \label{eq:H}
H(r)=h_c\bigg(\cfrac{r}{r_0}\bigg)^\psi r,
\end{equation} 
where \mbox{$h_c=0.125, \psi=0.05$};
for dust \mbox{$H_d(r)=\chi H(r)$}, where \mbox{$\chi=0.2$}~\citep{2019MNRAS.486.4638U}. The disk is assumed to be vertically isothermal. The temperature dependence on the distance to the star did not change over time and was set in such a way that the relation~(\ref{eq:H}).

\section{METHOD}\label{sec:method}
Gas-dynamic calculations were carried out using the SPH method, which is the basis of the cosmological code {\tt GADGET-2}\footnote{\url{https://wwwmpa.mpa-garching.mpg.de/gadget/}} \citep{2001NewA....6...79S,2005MNRAS.364.1105S}, modified for protoplanetary disks~\citep{2016Ap.....59..449D}. The calculations involved \mbox{$5\times10^5$} particles with gas properties and \mbox{$5\times10^5$} dust particles. The implementation of the scheme describing the interaction of gas and dust \citep{1995CoPhC..87..225M}, is described in  \citet{10.1007/978-3-031-49435-2_14}. The viscosity and self-gravity of the disk were taken into account.

In the calculations, the transfer of angular momentum in the disk was carried out by introducing numerical viscosity, for which the following parameters were set: \mbox{$\alpha_{\rm SPH}=0.2$, $\beta_{\rm SPH}=0$}. Turbulent viscosity in this case is defined as \mbox{$\nu=0.1\,c_s\,h\,\alpha_{\rm SPH}$} \citep{2012JCoPh.231..759P}, where $c_s$~ is the speed of sound determined within a sphere bounded by the smoothing length $h$ in the SPH method. Thus, the Shakura--Sunyaev viscosity parameter $\alpha_{\rm SS}$ $\alpha_{\rm SS}$~\citep{1973A&A....24..337S}  is not constant and is related to \mbox{$\alpha_{\rm SPH}$} by the ratio
\mbox{$\alpha_{\rm SS}=0.1\alpha_{\rm SPH}\,h/H(r)$}. 
In an unperturbed disk, the value of \mbox{$\alpha_{\rm SS}$} varies between 0.006 and 0.0001.

 For three-dimensional calculations of radiation transfer, the code  {\tt RADMC-3D}\footnote{\url{https://www.ita.uni-heidelberg.de/dullemond/software/radmc-3d/}} was used \citep{2012ascl.soft02015D}. The computational domain was divided into \mbox{$200\times30\times90$} cells in spherical coordinates (\mbox{$R,\theta,\phi$}), within which the average density of SPH particles of gas and, separately, dust was determined. It was assumed that the mass of fine dust (0.1\;$\mu$m) is \mbox{$4.9\times 10^{-5}\,M_\odot$}, and the total mass of coarse dust (1\;mm) is \mbox{$2.3\times 10^{-4}\,M_\odot$}~\citep{2019MNRAS.486.4638U}. $10^9$ photons were involved in the calculations of direct and scattered radiation. The dust opacity for magnesium-iron silicates \citep{1995A&A...300..503D} was calculated using the Mie theory \citep{1908AnP...330..377M}, using the code included in the {\tt RADMC-3D} package 
 \citep{1998asls.book.....B}. It was assumed that the disk is inclined by an angle \mbox{$i=35^\circ$}, the position angle $PA=55^\circ$, and the object is located at a distance of \mbox{$d=162$\;pc}~\citep{2019MNRAS.486.4638U}. 

The calculated radiation fluxes were used to simulate images that could potentially be obtained from observations with the ALMA radio interferometer. The simulation was carried out using a simulator {\tt CASA 6.5}\footnote{\url{https://casa.nrao.edu/}} \citep{2012ASPC..461..849P}. The calculations were performed at a wavelength of 1.3\;mm (Band\;6) for the CQ\,Tau position \mbox{$\alpha= 05^{\rm h} 35^{\rm m} 58\,.\!\!^{\rm s}46712$}, \mbox{$\delta = +24^\circ 44^\prime 54\,.\!\!^{\prime\prime}0864$}, the observer's bandwidth in the continuum is 6.8~GHz, the exposure time is about \mbox{$1^{\rm h}$} (corresponds to observations of 2017.1.01404.S, PI: L.~Testi). The antenna configuration, where the beam size was about $0\,.\!\!^{\prime\prime}15$, corresponded to Cycle~5 (5.7 of the available {\tt CASA} configurations). Thermal noise was added using the {\tt tsys-atm} option of the {\tt CASA} package, with the precipitable water vapor \mbox{$PWV = 0.6$}.

Calculations of the long-term dynamics of massless particles in the vicinity of a binary system to search for a stable orbit of a circumbinary planet were performed using the Bulirsh-Stoer algorithm ~\citep{1980.book.....S,1992nrca.book.....P}. The implementation of the method is described in detail in~\citet{2022A&C....4100635D}.

\section{RESULTS} \label{sec:res}

\subsection{Binary system}
\begin{figure*}[ht!]
\includegraphics[width=0.9\textwidth]{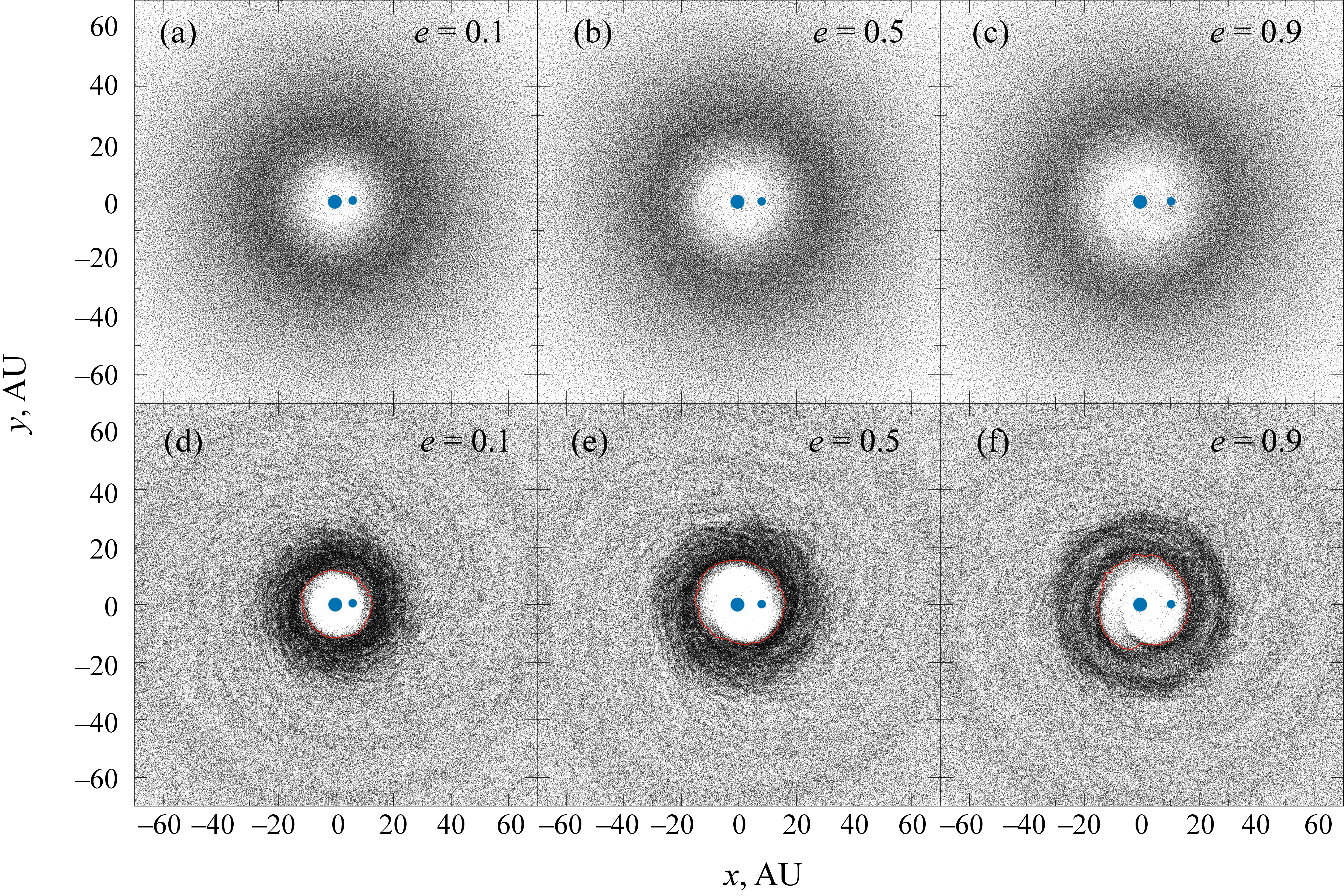}
\caption{ 
Projection of the positions of gas (panels (a)--(c)) and dust (panels (d)--(f)) particles onto the disk plane in a binary system model with companion mass $M_2=0.1$ and eccentricity \mbox{$e=0.1$} (panels (a), (d)), \mbox{$e=0.5$} (panels (b), (e)) and \mbox{$e=0.9$} (panels (c), (f)) at time \mbox{$T=1000$\;yr\;$=100\,P$}.
The red line shows the size of the cavity from Fig.~\ref{bgap} for the corresponding model.}\label{binaryE}
\end{figure*}

\begin{figure*}[ht!]
\includegraphics[width=0.81\textwidth]{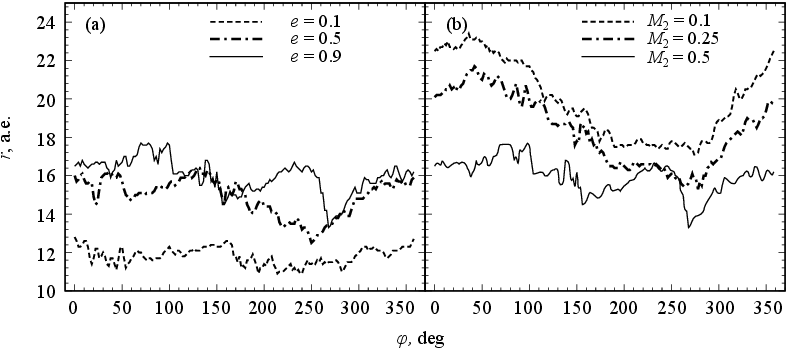}
\caption{Cavity size as a function of azimuthal angle $\phi$. Panel (a) shows the case with \mbox{$M_2=0.1\,M_\odot$} and three eccentricities: \mbox{$e=0.1$} (dashed line), \mbox{$e=0.5$} (dash-dot line), and \mbox{$e=0.9$} (solid line). Panel (b) shows the case with the eccentricity \mbox{$e=0.9$} for all models, where the companion mass varies: \mbox{$M_2=0.1\,M_\odot$} (solid line), \mbox{$M_2=0.25\,M_\odot$} (dash-dot line), and \mbox{$M_2=0.5\,M_\odot$} (dashed line). }\label{bgap}
\end{figure*}

Calculations were made of the dynamics of gas and dust particles in the gravitational field of a star (\mbox{$M_1=M_\odot$}) with a companion in an elongated orbit with a period of \mbox{$P=10$\;years} over $100$ orbital periods. A model with a companion mass of \mbox{$M_2=0.1\,M_\odot$} was considered, in which the eccentricity was varied: \mbox{$e=0.1,0.5,0.9$}. Calculations showed that the companion's orbital motion leads to the formation of a region of low gas density and a cavity in the distribution of dust particles in the central part of the disk (Fig.~\ref{binaryE}). The size of the cavity increases with increasing eccentricity, and its shape becomes asymmetric relative to the center of mass of the system (Fig.~\ref{bgap}a). However, even in the case of extreme eccentricity \mbox{$e=0.9$} the cavity size is 1.4~times smaller than that observed in CQ~Tau.

The next class of models considered the case of a highly elongated orbit with \mbox{$e=0.9$}, while varying the mass of the companion \mbox{$M_2=0.1,0.25,0.5\,M_\odot$}. In the presence of such a companion with a mass \mbox{$M_2=0.5\,M_\odot$}, the maximum velocity of the star at periastron reaches approximately \mbox{$18.1$\;km\,c$^{-1}$}, taking into account the disk inclination angle \mbox{$i=35^\circ$}, the radial velocity (\mbox{$v \sin i$}) reaches the value \mbox{10.4\;km\,c$^{-1}$}. This value is within in the kinematic local standard of rest frame (LSRK): \mbox{$15\pm2$\;km\,c$^{-1}$}~\citep{2022MNRAS.515.6109H}.

\begin{figure*}[ht!] \vspace{-2mm}
\includegraphics[width=0.8\textwidth]{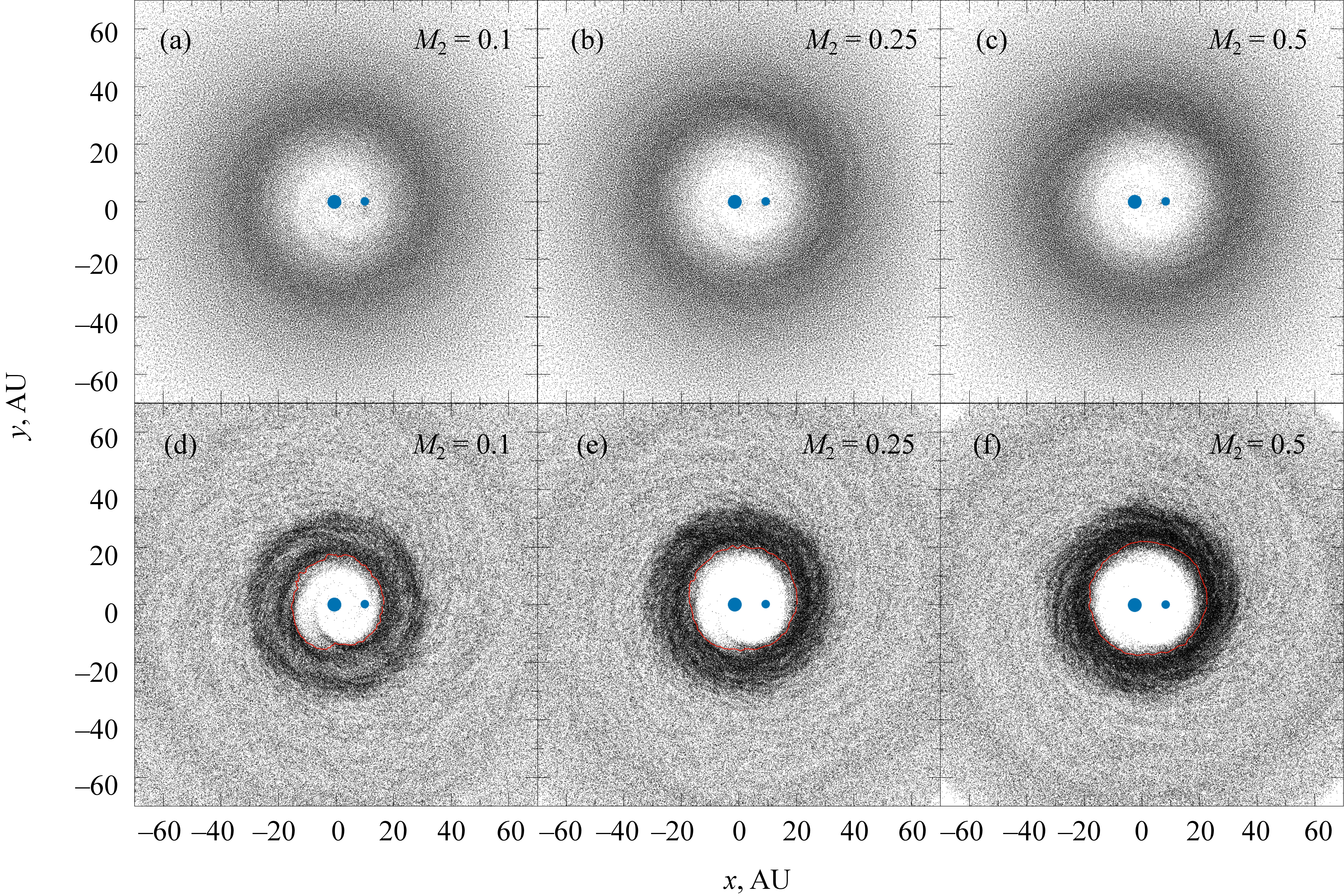}
\caption{\normalsize Same as in Fig.~\ref{binaryE} in the binary system model with eccentricity \mbox{$e=0.9$} and companion mass \mbox{$M_2=0.1\,M_\odot$} (panels (a), (d)), \mbox{$M_2=0.25\,M_\odot$} (panels (b), (e)) and \mbox{$M_2=0.5\,M_\odot$} (panels (c), (f)) at time $T=1000$\;years\;$=100\,P$. }\label{binaryM}
\end{figure*}

Calculations have shown that with an increase in the mass of the companion, the size and asymmetry of the cavity increases (Fig.~\ref{bgap}b). This is also associated with an increase in the semimajor axis of the binary system, which is \mbox{$a(0.1\,M_\odot)\approx 5.61$\;AU}, \mbox{$a(0.25\,M_\odot)\approx 5.77$~AU}, \mbox{$a(0.5M_\odot)\approx 6.01$\;AU} In the case of the most massive companion \mbox{$M_2=0.5\,M_\odot$}, the cavity size can reach 24\;AU In Fig.~\ref{binaryM} it is evident that with an increase in the mass of the companion, the proportion of gas penetrating into the cavity decreases. Thus, in the case of the companion \mbox{$M_2=0.1\,M_\odot$} after 100 of its revolutions, 4.7\% of the initial gas is retained inside the major semi-axis, with \mbox{$M_2=0.25$~is $2.1\%$} and with \mbox{$M_2=0.5$~is $1.5$\%}
 
Despite the presence of gas, large dust particles are subject to the influence of resonances. Therefore, with an increase in the mass of the companion, the co-orbital chaotic zone expands, within which the resonances of the mean motions are concentrated ~\citep{2015ApJ...799...41M,2020AstL...46..774D}. In Fig.~\ref{binaryE}, \ref{binaryM} it is seen that a dense dust ring is formed near the boundary of the chaotic zone. Its formation is apparently connected with the fact that dust particles, carried away by the gas, move toward the center of the system, and the presence of the chaotic zone prevents their further movement, which in turn leads to the accumulation of dust at its boundary.

\begin{figure*}[ht!]
\includegraphics[width=0.9\textwidth]{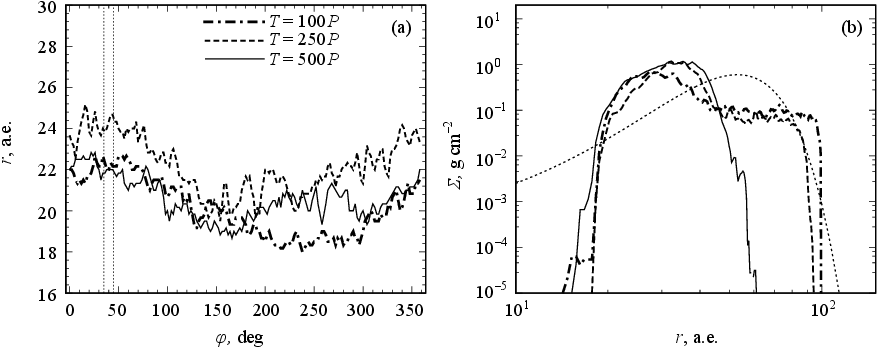}
\caption{\normalsize Dependence of the cavity size on the azimuthal angle $\phi$ for three times $100\,P$ (dash-dot line), $250\,P$ (dashed line) and $500\,P$ (solid line) for the model \mbox{$M_2=0.5\,M_\odot$}, \mbox{$e=0.9$} (a). Dependence of the surface density of large dust on the distance from the center of mass along the direction \mbox{$\phi\approx 40^\circ$} (shown in panel (a) by vertical dashed lines) for the same times (b). The dashed line describes the dependence~(\ref{Eq:sigmaObs}).}\label{TGap}
\end{figure*}

For the model with parameters \mbox{$M_2=0.5\,M_\odot$} and \mbox{$e=0.9$}, the long-term dynamics of dust and gas were calculated up to \mbox{$T=500\,P=5000$\;years}. Fig.~\ref{TGap}a shows the cavity size as a function of the azimuthal angle. It is evident that the cavity size changes slightly with time, but the shape of the dependence is preserved. The azimuthal direction \mbox{$\phi=40^\circ\pm 5^\circ$} was chosen near the maximum of this dependence, within which the dependence of the surface dust density on the distance from the center of mass was obtained. Fig.~\ref{TGap}b shows that over time, dust from the periphery of the disk shifts toward the center of the system and accumulates in a ring-shaped structure. In this case, the shape of the dependence of the surface density on the distance can be approximated by a Gaussian distribution.

In~\citet{2019MNRAS.486.4638U}, it was shown that the distribution of dust in the ring-shaped structure observed in the millimeter range corresponds to a Gaussian distribution:
\begin{equation}
\Sigma_{\rm obs}(r)=\Sigma_0\exp{\frac{-(r-r_0)^2}{2\sigma^2}}.
\label{Eq:sigmaObs}
\end{equation} 
The observations are best described with the parameters \mbox{$\Sigma_0=0.6$\;g\,cm$^{-2}$}, \mbox{$r_0=53$\;AU}, \mbox{$\sigma=13$} (the dotted line in Fig.~\ref{TGap}). Our calculations showed that in the case of a companion with a mass \mbox{$M_2=0.5\,M_\odot$} on a highly elongated orbit, the maximum in the dust distribution is reached at a distance of \mbox{$r\approx 35$\;AU}, which is noticeably closer than observed. Thus, the mass of the companion required to reproduce the observational characteristics of CQ\,Tau must be greater than $0.5\,M_\odot$. Such a massive companion would probably have already revealed itself during spectral observations. Therefore, another mechanism for the formation of a dust ring at a greater distance from the star is needed.

Note that the size of the cavity in the dust distribution even in the case of a companion \mbox{$M_2=0.1\,M_\odot$} with eccentricities \mbox{$e=0.5$} corresponds to the size of the inner region of the disk filled with matter in the model considered in~\citet{2019MNRAS.486.4638U}. Therefore, the model of a binary system with a planet can be one of the reasons for the formation of a cavity of approximately $25$\;AU and a suitable distribution of millimeter dust in the case of CQ\,Tau.

\subsection{Binary system with a planet}

It is impossible to prove the regularity of the orbit of a low-mass body in the restricted three-body problem by numerical methods~\citep[see, e.g.,][]{1999ssd..book.....M}. However, it is possible to estimate the regularity of the orbit over a limited time interval, which in our problem can be taken to be equal to the lifetime of the disk of $10^6$\;years~\citep{2001ApJ...553L.153H,2011ARA&A..49...67W}.

To determine the stable orbit of the planet, the dynamics of massless particles in the vicinity of a star with a low-mass companion (\mbox{$M_2=0.1,0.25\,M_\odot$}) on an elongated orbit (\mbox{$e=0.1,0.3,0.5$}) were calculated. The simulation involved $10^4$\;particles, the dynamics of which were considered over an interval of $10^6$\;years within the framework of the planar restricted three-body problem. The orbital pericenter argument ($\omega_p$) of massless particles and the true anomaly ($f_p$) were set randomly within $[0,2\pi]$, the eccentricity varied from 0.0\:\! to \:\!0.9, the semi-major axis was equal to \mbox{$a_{p}=20$\;AU} in three models, and varied from 25\:\! to 35\;AU in the others. Particles approaching the components of a binary system at a distance less than 1\%{} from the corresponding Hill radius,
\begin{equation*}
H_{M_1}=\sqrt[3]{\cfrac{M_1}{3M_2}\,a},\;\; H_{M_2}=\sqrt[3]{\cfrac{M_2}{3M_1}\,a},
\end{equation*}
and those that went to a distance greater than 100\;AU from the center of the system were considered to have left the system.
 
\renewcommand{\baselinestretch}{0.85}
\begin{table}
\caption {Parameters of models of binary systems with a planet}\label{models} 
\medskip
\begin{tabular}{l|c|c|c|c|c|c|c}
\hline
$N$&$M_2$, $M_\odot$&$a$, AU&$e$& $a_p$, AU&$e_p$&$\omega_p$, deg &$f_p$, deg \\ 
\hline
1   &  $0.1$ & $5.62$ & $0.1$ & $20$ & $0.32$ & $140.94$ & $10.29$\\
2   &  $0.1$ & $5.62$ & $0.3$ & $20$ & $0.29$ & $147.35$ & $12.85$\\
3   &  $0.1$ & $5.62$ & $0.5$ & $20$ & $0.30$ & $192.84$ & $51.35$\\\hline
4   &  $0.1$ & $5.62$ & $0.5$ & $26.64$ & $0.32$ & $173.78$  & $235.19$\\
5   &  $0.1$ & $5.62$ & $0.5$ & $29.79$ & $0.14$ & $190.69$ & $218.65$\\
6  &  $0.1$ & $5.62$ & $0.5$ & $30.68$ & $0.45$ & $201.28$ & $32.17$\\
7  &  $0.1$ & $5.62$ & $0.5$ & $34.42$ & $0.44$ & $184.18$ & $269.53$\\\hline
8  &  $0.25$ & $5.77$ & $0.5$ & $31.79$ & $0.19$ & $158.26$ & $130.23$\\
9  &  $0.25$ & $5.77$ & $0.5$ & $29.73$ & $0.34$ & $206.72$ & $335.48$\\
10  &  $0.25$ & $5.77$ & $0.5$ & $32.40$ & $0.24$ & $161.40$ & $263.63$\\
11  &  $0.25$ & $5.77$ & $0.5$ & $30.02$ & $0.21$ & $178.46$ & $182.73$\\
\hline
\end{tabular}
\end{table} \renewcommand{\baselinestretch}{1}

Calculations have shown that the number of particles remaining in the system at the end of the calculations decreases with increasing eccentricity. Thus, in the case of models with a constant semi-major axis and a companion mass \mbox{$M_2=0.1\,M_\odot$}, the initial eccentricity of the surviving orbits for all models is less than 0.32. In the case of a binary system eccentricity \mbox{$e=0.1$}, 34.8\%{} particles remain in the system, in the case of \mbox{$e=0.3$}~are 24.0\%{} and for \mbox{$e=0.5$}~are 15.8\%{} particles. With increasing companion mass, the number of ``surviving'' particles also decreases. In models where the semi-major axis of the particles varied \mbox{$a_p=25$--$35$\;AU}, in the case of \mbox{$M_2=0.1\,M_\odot$}, 39.3\%{} were preserved in the system, and in the case of \mbox{$M_2=0.25\,M_\odot$}~are 33.2\%{}. The eccentricities of the preserved orbits in the first case are larger on average than in the second, but do not exceed 0.5 (with the exception of a few cases where \mbox{$e_p>0.8$}).

\begin{figure*}[ht!]
\includegraphics[width=1.0\textwidth]{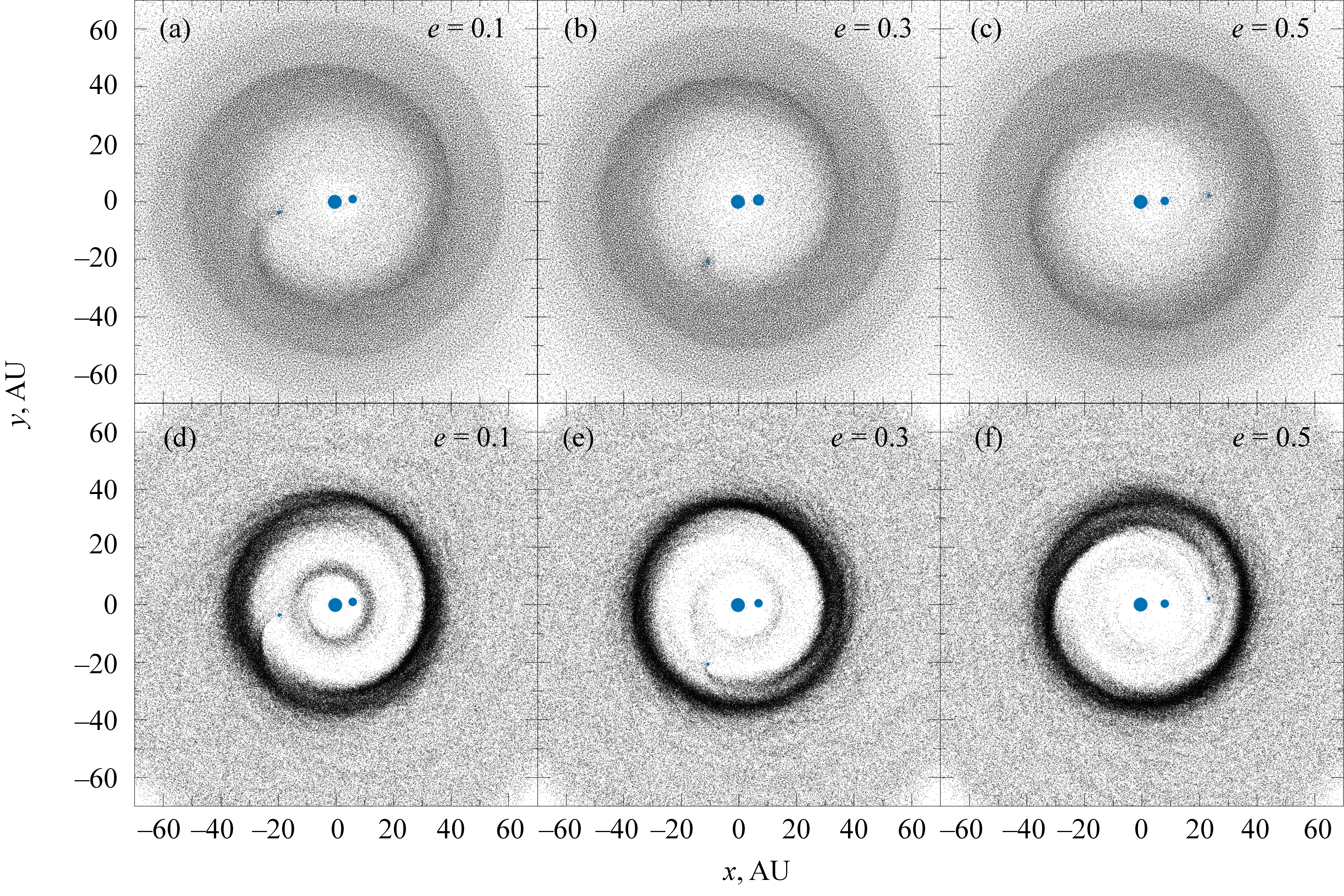}
\caption{\normalsize Same as Fig.~\ref{binaryE} for the binary system model with a planet. Model~1 is shown on the left, Model~2 in the center and Model~3 on the right (from Table~\ref{models} at time \mbox{$T=200\,P\approx 30\,P_p$}). The eccentricity of the binary system is indicated in the upper right corner. }\label{planet}
\end{figure*}

To calculate the dynamics of dust and gas in a binary system with a planet, a number of orbits were chosen whose eccentricities and semi-major axes change little during the calculations. The mass of the main component was set equal to \mbox{$M_1=1.67\,M_\odot$}, and the planet's mass is \mbox{$m_p=9\,M_{\rm Jup}$}, the period of the binary system~is $P=10$\;years for all models. The remaining parameters of the models are given in Table~\ref{models}.

The choice of models with a noticeable eccentricity of the planet is due to the fact that when calculating the dynamics of massless particles in systems of single and binary stars with planets, a co-orbital ring with the planet is stabilized if the binary system presence at the center ~\citep{2016MNRAS.463L..22D, 2018SoSyR..52..180D}.
At the same time, the co-orbital structure can be destroyed over time in the case of significant eccentricity of the binary system and the planet~\citep{2018AstL...44..119D}. %[Demidova, Shevchenko, 2018].
The presence of a co-orbital ring reduces the apparent size of the cavity free of matter. In addition, the greater eccentricity of the planet contributes to an increase in the width of the cavity. 

\begin{figure*}[]
\includegraphics[width=0.81\textwidth]{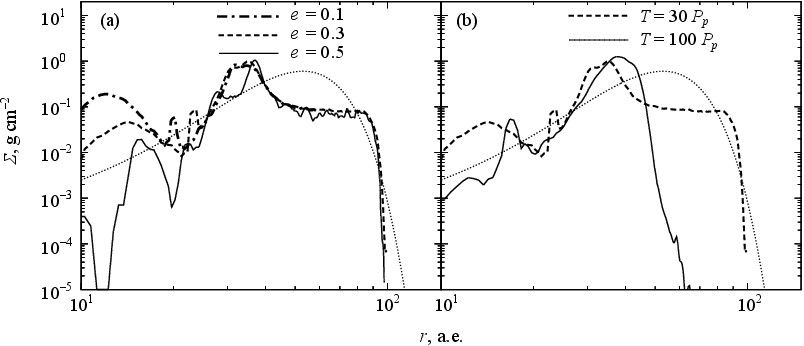}
\caption{\normalsize Azimuthally averaged surface density as a function of distance from the center of mass of the binary system. On the left are the results for Model~1 (dash-dotted line), Model~2 (dashed line), and Model~3 (solid line) from Table~\ref{models} at time \mbox{$T\approx30\,P_p$}. On the right are the results for Model~2 at times \mbox{$T\approx30\,P_p$} (dashed line) and \mbox{$T\approx100\,P_p$} (solid line). The dotted line describes the dependence~(\ref{Eq:sigmaObs}). }\label{planetA20}
\end{figure*}

Fig.~\ref{planet} shows the results of calculations for the models with \mbox{$a_p=20$\;AU} and varying $e$ (Models 1--3). It is evident that, just as in the case of a binary system without a planet, in the model with a planet dust accumulates near the chaotic zone, in this case associated with the planet. The amount of gas penetrating into the central parts of the system corresponds to that for the case of a binary system without a planet. At small $e$ a dust ring is preserved between the orbit of the binary star and the planet; with increasing \mbox{$e\ge 0.3$} this structure dissipates. The position of the dust structure on the outer boundary of the chaotic zone of the planet's orbit does not depend on the eccentricity of the binary star (Fig.~\ref{planetA20}a). In addition, as can be seen in Fig.~\ref{planetA20}, dust shifts toward the center of the system over time and accumulates in a ring-shaped density perturbation. For this class of models, the maximum in the surface density distribution in the dust ring is located at a distance of 37.5\;AU at the time \mbox{$T=6750$\;yr~$\approx100\,P_p$}. Thus, the structure is located significantly closer than the observed one.

\begin{figure*}[ht!]
\includegraphics[width=0.9\textwidth]{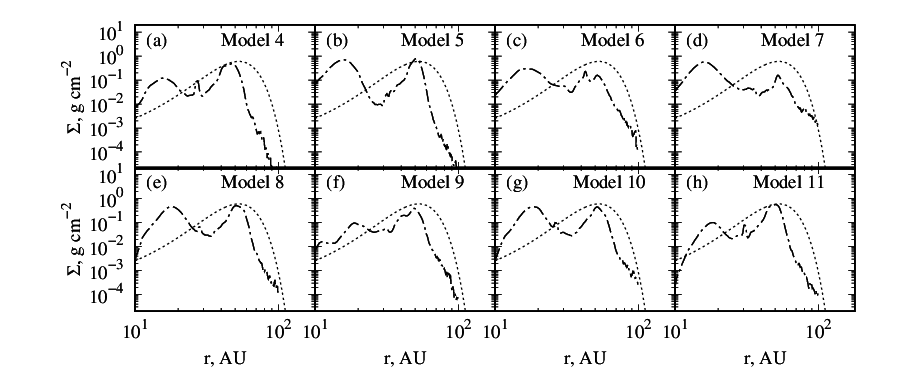}
\caption{\normalsize Same as in Fig.~\ref{planetA20} at time \mbox{$T=100P_p$} (dash-dotted line) for Models 4--11 (numbers are indicated at the top of each graph). The dotted line describes the dependence~(\ref{Eq:sigmaObs}).}\label{planetSigma}
\end{figure*}

In the next class of models (Models 4--7), calculations were performed with the parameters \mbox{$M_2=0.1$} and \mbox{$e=0.5$}, \mbox{$a_p=25$--$35$\;AU}, \mbox{$e_p=0.14$--$0.45$}. The top panels of Fig.~\ref{planetSigma} show the dependence of the azimuthally averaged surface density on the distance to the center of mass. It is evident that in all the models considered, the dust structure is preserved between the orbit of the binary star and the planet, and the second ring can form on the outer boundary of the planet's chaotic zone. However, with an increase in $e_p$ to the value \mbox{$e_p\sim 0.45$}, the outer dust ring is weakly expressed. It should be noted that in Model~5 (\mbox{$a_p\approx 30$\;AU}) the maximum in the distribution of the surface density of dust is reached at a distance of $51$\;AU, which is close to the observed value.

\begin{figure*}[ht!]
\includegraphics[width=0.99\textwidth]{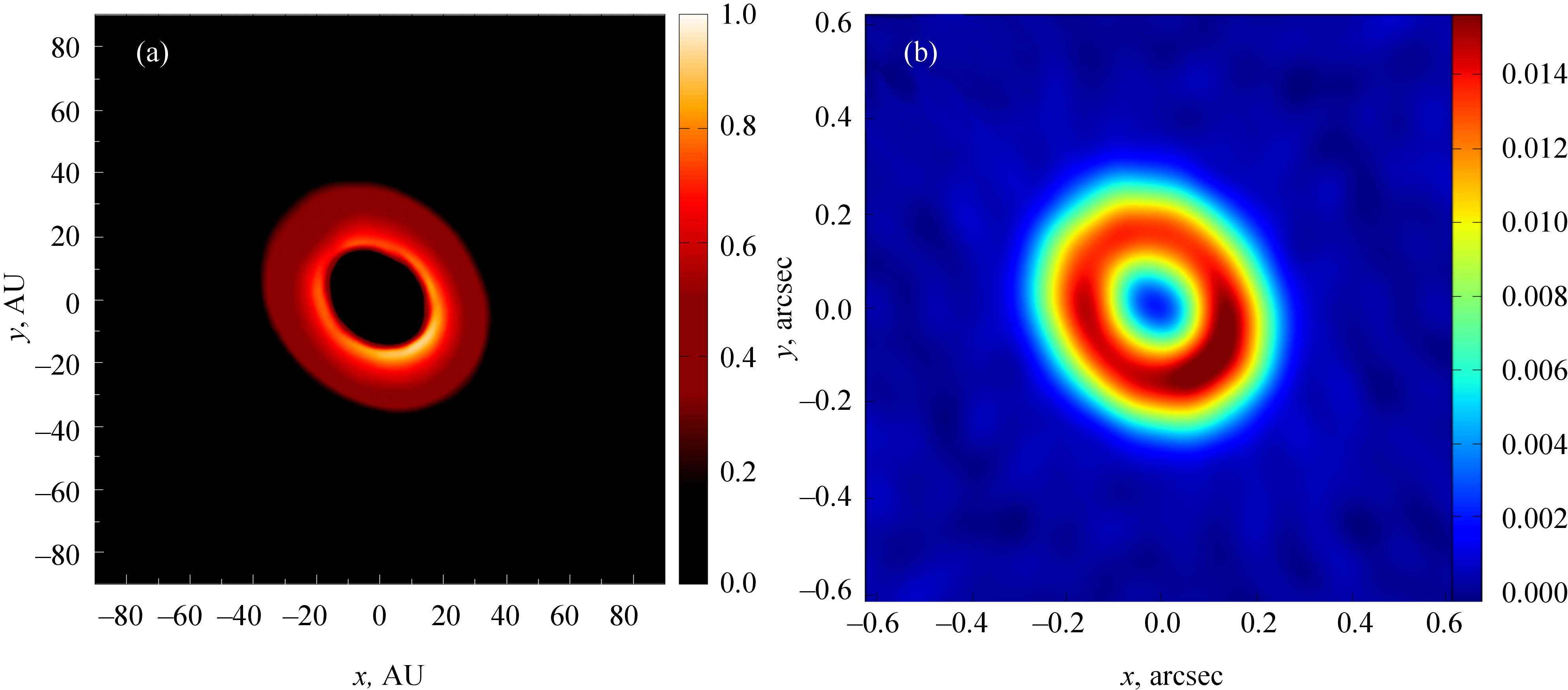}
\caption{Panel (a) shows the radiative flux in a binary system model without a planet. The color scale is in \mbox{$10^{-12}$~erg\,c$^{-1}$\,cm$^{-2}$\,Hz$^{-1}$\,sr$^{-1}$}. Panel (b)~is a theoretical image at a wavelength of 1.3\;mm. The color scale is in Jy\,beam$^{-1}$. The model parameters are \mbox{$M_2=0.5\,M_\odot$}, \mbox{$P=10$\;yr}, \mbox{$e=0.9$}, at time \mbox{$T=500\,P$.}}\label{bimage}
\end{figure*}
\begin{figure*}[ht!] \vspace{-3mm}
\includegraphics[width=0.99\textwidth]{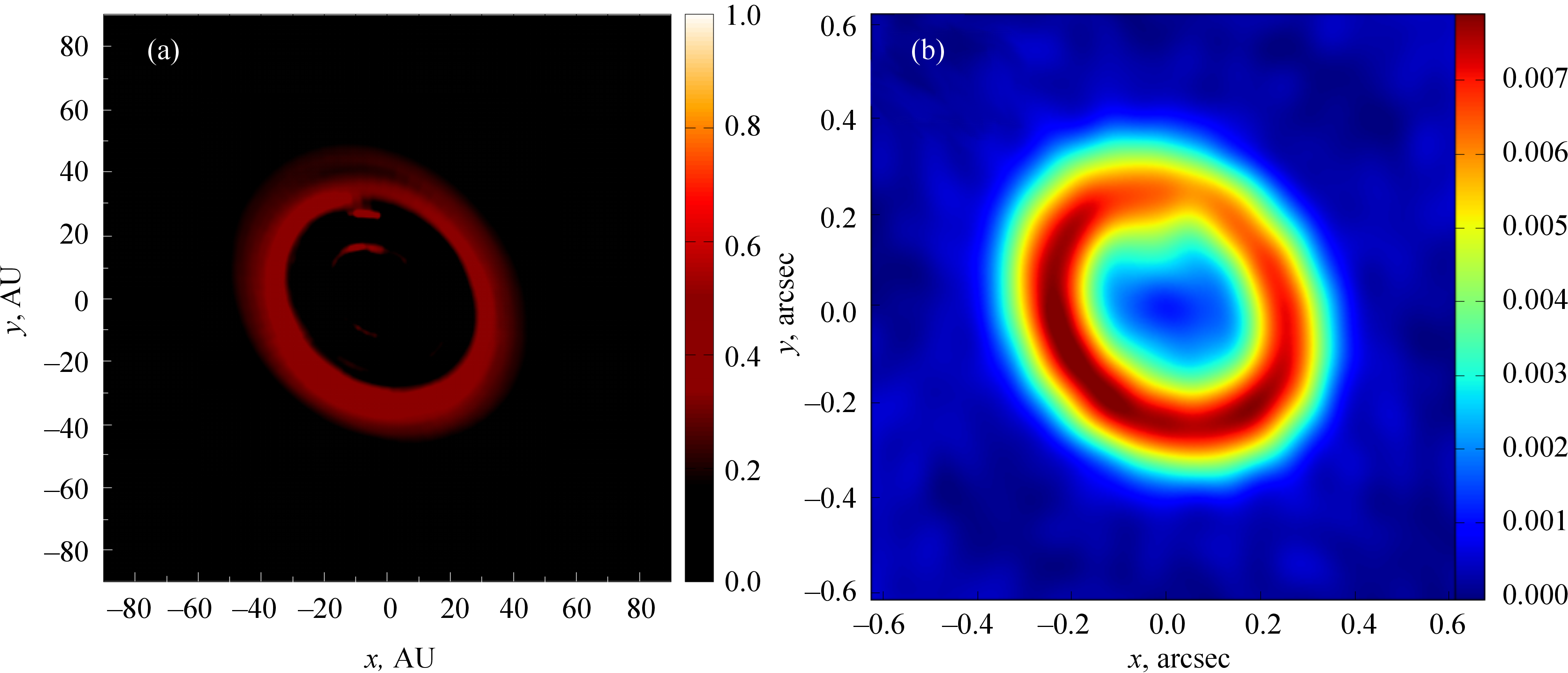}
\caption{\normalsize Same as in Fig.~\ref{bimage} for a binary system with a planet. Model parameters \mbox{$M_2=0.25\,M_\odot$}, $P=10$\;yr, $e=0.5$, $m_p=9\,M_{\rm Jup}$, $a_p=30.02$\;AU, $e_p\approx 0.21$ at time $T\approx 1000\,P$.}\label{pimage}
\end{figure*}

Therefore, in the next class of models, the mass of the companion was increased to $M_2=0.25\,M_\odot$ and orbits near $a_p=30$\;AU were considered. The bottom panels of Fig.~\ref{planetSigma} show that at $a_p>30$\;AU (Models 8, 10) and $e_p\approx 0.2$, a dust ring is preserved between the orbits. At $a\approx 30$\;AU, in Model~9 with a large $e_p$, large dust between the orbits is removed more intensively than in Model~11. Moreover, in the first case, the maximum of the surface dust density distribution lies at a distance of $49.5$\;AU, and in the second~is $50.5$\;AU. In addition, in the second case, the outer ring-shaped structure is more symmetrical due to the lower eccentricity of the planet.

\subsection{Disc images}
Based on gas-dynamic models, the dust temperature and the radiation flux from the protoplanetary disk at a wavelength of 1.3 mm were calculated. Then theoretical images of the disk were constructed, which can be obtained using the ALMA radio interferometer.

Fig.~\ref{bimage} shows a case of a binary system with a companion $M_2=0.5\,M_\odot$ and $e=0.9$. In the calculations it was assumed that the companion is a class M star, which has a radius $R_2=0.5\,R_\odot$ and a temperature of 3600\;K. It is seen that the image has a ring-shaped form. The size of the central cavity is $0\,.\!\!^{\prime\prime}11$. At the same time, the maximum intensity reaches 17~mJy\,beam$^{-1}$.

For a binary system with a planet, the disk image is constructed for Model 11 (Fig.~\ref{pimage}). It was assumed that the companion has the following characteristics: $M_2=0.25\,M_\odot$, $R_2=0.3\,R_\odot$, $T_2=3000$~K, $e=0.5$. In this case, the central cavity has a size of $0\,.\!\!^{\prime\prime}27$, which is in good agreement with the data obtained from observations of~\citep{2019MNRAS.486.4638U}. The maximum flux intensity is 8.1\;mJy\,beam$^{-1}$. The image shows that the ring-shaped dust structure is distorted, since it has the shape of a tightly twisted spiral. Probably, the eccentricity of the planet should be smaller to maintain an undistorted ring-shaped structure. Thus, the binary system model with a planet better describes the observational data.

\section{CONCLUSION}\label{sec:concl}

Calculations have shown that millimeter-sized dust particles tend to move by gas in the direction of the system's center of mass, but cannot overcome the barrier at the boundary of the chaotic zone formed by the condensation of resonances with the period of the binary system or planet. Therefore, dust can accumulate at the outer boundary of the chaotic zone, forming a ring-shaped compaction. Gas with fine dust can penetrate into the inner regions of the system. This agrees well with the fact that the size of the cavity determined by the lines of CO molecules is smaller than in the continuum ($\lambda=1.3$\;mm).

The calculation results showed that only a massive companion on a highly eccentric orbit is capable of clearing the cavity observed in the CQ\,Tau image. It should be noted that the cavity size depends not only on the companion-to-star mass ratio and eccentricity, but also on the disk half-thickness. In \citet{2020MNRAS.498.2936H} it was shown that at $H/r=0.01$ the cavity size for a binary star with mass ratio $q=0.1$ and eccentricity $e=0.8$ can reach $5a$ (about 28\;AU at $a=5.61$\;AU) after 1000~revolutions. However, in \citet{2019MNRAS.486.4638U} it is shown that the best description of the observations is given by the half-thickness $$H/r=0.125\bigg(\cfrac{r}{r_0}\bigg)^{0.05},$$ which is significantly larger than 0.01. At the same time, the dust ring in this model is significantly closer than determined in~\citet{2019MNRAS.486.4638U}.

Another mechanism influencing the dust distribution is the enlargement and settling of dust grains toward the disk plane during its evolution. This leads to a decrease in the amount of submillimeter and millimeter dust in the central parts of the disk.
As shown in~\citet{2005A&A...434..971D,2008A&A...480..859B}, this process can remove such dust from the inner parts of the disk on a ``short'' time interval (about $10^4$ years). At the same time, micron dust may be present due to collisions of larger dust grains and planetesimals \citep{2009A&A...503L...5B}.

Studies of planetesimal dynamics in the debris disk of a binary system have shown that in the case of a non-circular orbit of a binary star, their eccentricities are pumped \citep{2004ApJ...609.1065M,2012ApJ...754L..16P,2012ApJ...752...71M}.
This can lead to an increase in the number of intersecting orbits of planetesimals and their more efficient collisions. In addition, it was shown that in the inner part ($\leq 4 a_b$) of the system, planetesimals can transition to hyperbolic orbits. However, the effect of eccentricity pumping can be reduced by the influence of disk self-gravity~\citep{2023ApJ...954..100S}. It should also be noted that when planetesimals and large dust particles move in orbits with significant eccentricity, the time spent in the orbital apoastron significantly exceeds the time of periastron passage, which can lead to a visible decrease in the density of such particles in the inner part of the system.
In~\citet{2001ApJ...557..990T}, the possibility of dust migration near stars of early spectral classes under the influence of stellar radiation was considered. Fine dust (less than 1\;mm) tends to migrate outward from the disk and concentrate on its periphery, forming a ring-shaped structure. Whereas larger dust (more than 1\;mm), on the contrary, shifts toward the central part of the disk.

Another reason for the formation of a cavity in the gas-dust disk of CQ\,Tau may be a planet orbiting a binary system. As calculations have shown, even in the vicinity of a binary system with an elongated orbit $e=0.5$, it is possible for a planet to exist on an orbit that is stable over the lifetime of the disk. The issue of planet stability in binary systems with the planets Kepler-16b, Kepler-34b, and Kepler-35b was considered in~\citet{2013ApJ...769..152P}. It was shown that planets discovered in these systems may be located inside stable resonant cells, between chaotic regions.

To date, 21 exoplanets have been discovered orbiting a binary system\footnote{according to the website \url{https://exoplanet.eu}}. Two objects have an orbital eccentricity of the binary system: $e>0.5$ (Kepler-34, Kepler1660). In the case of Kepler-34, the planet eccentricity is $e_p=0.182$, and the maximum eccentricity was detected for the planet DP\,Leo\,b: $e_p=0.390$. The minimum distance from the central binary is $a_p\approx 2a$. It is worth noting that the identified exoplanets orbit close binary pairs with a major semi-axis \mbox{$a<1$}\;AU (except for the object OGLE2023BLG0836, where \mbox{$a=1.88$}\;AU). This selection may be due to the selection effect of exoplanet detection methods.

Thus, the probability of the presence of a binary system with a low-mass companion and a common planet in the central cavity of CQ\,Tau is not zero. However, in order to determine the stability zones for a hypothetical planet, it is necessary to know the orbital eccentricity and the mass of the companion. Calculations have shown that the observations correspond to models in which the mass of the companion is \mbox{$M_2\ge0.25\,M_\odot$}, and the eccentricity is $e\ge0.5$, while the massive planet should be located at an approximate distance of 30\;AU and have a small eccentricity $e<0.2$.

It should be emphasized that the search for possibilities of detecting planets orbiting wider pairs (including those at the stage of formation in the protoplanetary disk) is an urgent task, the solution of which can be facilitated by further study of CQ\,Tau.

\section*{ACKNOWLEDGMENTS}
The author expresses gratitude to V.~P.~Grinin for scientific consultation on the object CQ\,Tau. Calculations were carried out using the resources of the Interdepartmental Supercomputer Center of the Russian Academy of Sciences, a branch of the Federal State Institution ``Research Institute for Systems Analysis of the Russian Academy of Sciences'' \citep{2019LG..40..1835}.
 \bibliography{biblio}{}

\begin{thebibliography}{43}
% this bibliography was produced with the style dinat.bst v2.5
\makeatletter
\newcommand{\dinatlabel}[1]%
{\ifNAT@numbers\else\NAT@biblabelnum{#1}\hspace{2\labelsep}\fi}
\makeatother
\expandafter\ifx\csname natexlab\endcsname\relax\def\natexlab#1{#1}\fi
\expandafter\ifx\csname url\endcsname\relax\def\url#1{\texttt{#1}}\fi

\bibitem[{Artymowicz} und {Lubow}(1994)]{1994ApJ...421..651A}
\dinatlabel{{Artymowicz} und {Lubow} 1994} \textsc{{Artymowicz}}, Pawel~;
  \textsc{{Lubow}}, Stephen~H.:
\newblock {Dynamics of Binary-Disk Interaction. I. Resonances and Disk Gap
  Sizes}.
\newblock In: \emph{\apj}
\newblock 421 (1994), Februar, S.~651

\bibitem[{Birnstiel} u.\,a.(2009){Birnstiel}, {Dullemond} und
  {Brauer}]{2009A&A...503L...5B}
\dinatlabel{{Birnstiel} u.\,a. 2009} \textsc{{Birnstiel}}, T.~;
  \textsc{{Dullemond}}, C.~P.~; \textsc{{Brauer}}, F.:
\newblock {Dust retention in protoplanetary disks}.
\newblock In: \emph{\aap}
\newblock 503 (2009), August, Nr.~1, S.~L5--L8

\bibitem[{Bohren} und {Huffman}(1998)]{1998asls.book.....B}
\dinatlabel{{Bohren} und {Huffman} 1998} \textsc{{Bohren}}, Craig~F.~;
  \textsc{{Huffman}}, Donald~R.:
\newblock \emph{{Absorption and Scattering of Light by Small Particles}}.
\newblock 1998

\bibitem[{Brauer} u.\,a.(2008){Brauer}, {Dullemond} und
  {Henning}]{2008A&A...480..859B}
\dinatlabel{{Brauer} u.\,a. 2008} \textsc{{Brauer}}, F.~; \textsc{{Dullemond}},
  C.~P.~; \textsc{{Henning}}, Th.:
\newblock {Coagulation, fragmentation and radial motion of solid particles in
  protoplanetary disks}.
\newblock In: \emph{\aap}
\newblock 480 (2008), M\^^b{a}rz, Nr.~3, S.~859--877

\bibitem[{Demidova}(2022)]{2022A&C....4100635D}
\dinatlabel{{Demidova} 2022} \textsc{{Demidova}}, T.:
\newblock {Bulirsh-Stoer algorithm in the planar restricted three-body
  problem}.
\newblock In: \emph{Astronomy and Computing}
\newblock 41 (2022), Oktober, S.~100635

\bibitem[{Demidova}(2016)]{2016Ap.....59..449D}
\dinatlabel{{Demidova} 2016} \textsc{{Demidova}}, T.~V.:
\newblock {Modelling the Gas Dynamics of Protoplanetary Disks by the SPH
  Method}.
\newblock In: \emph{Astrophysics}
\newblock 59 (2016), Dec, Nr.~4, S.~449--460

\bibitem[{Demidova}(2018)]{2018SoSyR..52..180D}
\dinatlabel{{Demidova} 2018} \textsc{{Demidova}}, T.~V.:
\newblock {``Horseshoe'' Structures in the Debris Disks of Planet-Hosting
  Binary Stars}.
\newblock In: \emph{Solar System Research}
\newblock 52 (2018), M\^^b{a}rz, Nr.~2, S.~180--188

\bibitem[{Demidova} und {Grinin}(2017)]{2017AstL...43..106D}
\dinatlabel{{Demidova} und {Grinin} 2017} \textsc{{Demidova}}, T.~V.~;
  \textsc{{Grinin}}, V.~P.:
\newblock {SPH simulations of structures in protoplanetary disks}.
\newblock In: \emph{Astronomy Letters}
\newblock 43 (2017), Februar, Nr.~2, S.~106--119

\bibitem[{Demidova} und {Shevchenko}(2018)]{2018AstL...44..119D}
\dinatlabel{{Demidova} und {Shevchenko} 2018} \textsc{{Demidova}}, T.~V.~;
  \textsc{{Shevchenko}}, I.~I.:
\newblock {Simulations of the Dynamics of the Debris Disks in the Systems
  Kepler-16, Kepler-34, and Kepler-35}.
\newblock In: \emph{Astronomy Letters}
\newblock 44 (2018), Februar, Nr.~2, S.~119--125

\bibitem[{Demidova} und {Shevchenko}(2020)]{2020AstL...46..774D}
\dinatlabel{{Demidova} und {Shevchenko} 2020} \textsc{{Demidova}}, T.~V.~;
  \textsc{{Shevchenko}}, I.~I.:
\newblock {Long-Term Dynamics of Planetesimals in Planetary Chaotic Zones}.
\newblock In: \emph{Astronomy Letters}
\newblock 46 (2020), November, Nr.~11, S.~774--782

\bibitem[Demidova u.\,a.(2023)Demidova, Savvateeva, Anoshin, Grigoryev und
  Stoyanovskaya]{10.1007/978-3-031-49435-2_14}
\dinatlabel{Demidova u.\,a. 2023} \textsc{Demidova}, Tatiana~;
  \textsc{Savvateeva}, Tatiana~; \textsc{Anoshin}, Sergey~; \textsc{Grigoryev},
  Vitaliy~; \textsc{Stoyanovskaya}, Olga:
\newblock Implementation of Dusty Gas Model Based on Fast and Implicit
  Particle-Mesh Approach SPH-IDIC in Open-Source Astrophysical Code GADGET-2.
\newblock In: \textsc{Voevodin}, Vladimir (Hrsg.)~; \textsc{Sobolev}, Sergey
  (Hrsg.)~; \textsc{Yakobovskiy}, Mikhail (Hrsg.)~; \textsc{Shagaliev}, Rashit
  (Hrsg.): \emph{Supercomputing}.
\newblock Cham~: Springer Nature Switzerland, 2023, S.~195--208. --
\newblock ISBN 978-3-031-49435-2

\bibitem[{Demidova} und {Shevchenko}(2016)]{2016MNRAS.463L..22D}
\dinatlabel{{Demidova} und {Shevchenko} 2016} \textsc{{Demidova}}, Tatiana~V.~;
  \textsc{{Shevchenko}}, Ivan~I.:
\newblock {Three-lane and multilane signatures of planets in planetesimal
  discs}.
\newblock In: \emph{\mnras}
\newblock 463 (2016), November, Nr.~1, S.~L22--L25

\bibitem[{Dorschner} u.\,a.(1995){Dorschner}, {Begemann}, {Henning}, {Jaeger}
  und {Mutschke}]{1995A&A...300..503D}
\dinatlabel{{Dorschner} u.\,a. 1995} \textsc{{Dorschner}}, J.~;
  \textsc{{Begemann}}, B.~; \textsc{{Henning}}, T.~; \textsc{{Jaeger}}, C.~;
  \textsc{{Mutschke}}, H.:
\newblock {Steps toward interstellar silicate mineralogy. II. Study of
  Mg-Fe-silicate glasses of variable composition.}
\newblock In: \emph{Astron. Astrophys.}
\newblock 300 (1995), Aug, S.~503

\bibitem[{Dullemond} und {Dominik}(2005)]{2005A&A...434..971D}
\dinatlabel{{Dullemond} und {Dominik} 2005} \textsc{{Dullemond}}, C.~P.~;
  \textsc{{Dominik}}, C.:
\newblock {Dust coagulation in protoplanetary disks: A rapid depletion of small
  grains}.
\newblock In: \emph{\aap}
\newblock 434 (2005), Mai, Nr.~3, S.~971--986

\bibitem[{Dullemond} u.\,a.(2012){Dullemond}, {Juhasz}, {Pohl}, {Sereshti},
  {Shetty}, {Peters}, {Commercon} und {Flock}]{2012ascl.soft02015D}
\dinatlabel{{Dullemond} u.\,a. 2012} \textsc{{Dullemond}}, C.~P.~;
  \textsc{{Juhasz}}, A.~; \textsc{{Pohl}}, A.~; \textsc{{Sereshti}}, F.~;
  \textsc{{Shetty}}, R.~; \textsc{{Peters}}, T.~; \textsc{{Commercon}}, B.~;
  \textsc{{Flock}}, M.:
\newblock \emph{{RADMC-3D: A multi-purpose radiative transfer tool}}.
\newblock Februar 2012

\bibitem[{Dutrey} u.\,a.(1994){Dutrey}, {Guilloteau} und
  {Simon}]{1994A&A...286..149D}
\dinatlabel{{Dutrey} u.\,a. 1994} \textsc{{Dutrey}}, A.~;
  \textsc{{Guilloteau}}, S.~; \textsc{{Simon}}, M.:
\newblock {Images of the GG Tauri rotating ring}.
\newblock In: \emph{Astron. Astrophys.}
\newblock 286 (1994), Juni, S.~149--159

\bibitem[{Grinin} u.\,a.(2023){Grinin}, {Tambovtseva}, {Barsunova} und
  {Shakhovskoy}]{2023Ap.....66..235G}
\dinatlabel{{Grinin} u.\,a. 2023} \textsc{{Grinin}}, V.~P.~;
  \textsc{{Tambovtseva}}, L.~V.~; \textsc{{Barsunova}}, O.~Y.~;
  \textsc{{Shakhovskoy}}, D.~N.:
\newblock {Photometric Activity of CQ Tau over a Time Interval of 125 Years}.
\newblock In: \emph{Astrophysics}
\newblock 66 (2023), Juni, Nr.~2, S.~235--241

\bibitem[{Haisch} u.\,a.(2001){Haisch}, {Lada} und {Lada}]{2001ApJ...553L.153H}
\dinatlabel{{Haisch} u.\,a. 2001} \textsc{{Haisch}}, Jr.~; \textsc{{Lada}},
  Elizabeth~A.~; \textsc{{Lada}}, Charles~J.:
\newblock {Disk Frequencies and Lifetimes in Young Clusters}.
\newblock In: \emph{\apjl}
\newblock 553 (2001), Juni, Nr.~2, S.~L153--L156

\bibitem[{Hammond} u.\,a.(2022){Hammond}, {Christiaens}, {Price},
  {Ubeira-Gabellini}, {Baird}, {Calcino}, {Benisty}, {Lodato}, {Testi},
  {Pinte}, {Toci} und {Fedele}]{2022MNRAS.515.6109H}
\dinatlabel{{Hammond} u.\,a. 2022} \textsc{{Hammond}}, Iain~;
  \textsc{{Christiaens}}, Valentin~; \textsc{{Price}}, Daniel~J.~;
  \textsc{{Ubeira-Gabellini}}, Maria~G.~; \textsc{{Baird}}, Jennifer~;
  \textsc{{Calcino}}, Josh~; \textsc{{Benisty}}, Myriam~; \textsc{{Lodato}},
  Giuseppe~; \textsc{{Testi}}, Leonardo~; \textsc{{Pinte}}, Christophe~;
  \textsc{{Toci}}, Claudia~; \textsc{{Fedele}}, Davide:
\newblock {External or internal companion exciting the spiral arms in CQ Tau?}
\newblock In: \emph{\mnras}
\newblock 515 (2022), Oktober, Nr.~4, S.~6109--6121

\bibitem[{Hirsh} u.\,a.(2020){Hirsh}, {Price}, {Gonzalez}, {Ubeira-Gabellini}
  und {Ragusa}]{2020MNRAS.498.2936H}
\dinatlabel{{Hirsh} u.\,a. 2020} \textsc{{Hirsh}}, Kieran~; \textsc{{Price}},
  Daniel~J.~; \textsc{{Gonzalez}}, Jean-Fran{\c{c}}ois~;
  \textsc{{Ubeira-Gabellini}}, M.~G.~; \textsc{{Ragusa}}, Enrico:
\newblock {On the cavity size in circumbinary discs}.
\newblock In: \emph{\mnras}
\newblock 498 (2020), Oktober, Nr.~2, S.~2936--2947

\bibitem[{Meschiari}(2012)]{2012ApJ...752...71M}
\dinatlabel{{Meschiari} 2012} \textsc{{Meschiari}}, Stefano:
\newblock {Circumbinary Planet Formation in the Kepler-16 System. I. N-body
  Simulations}.
\newblock In: \emph{\apj}
\newblock 752 (2012), Juni, Nr.~1, S.~71

\bibitem[{Mie}(1908)]{1908AnP...330..377M}
\dinatlabel{{Mie} 1908} \textsc{{Mie}}, Gustav:
\newblock {Beitr{\"a}ge zur Optik tr{\"u}ber Medien, speziell kolloidaler
  Metall{\"o}sungen}.
\newblock In: \emph{Annalen der Physik}
\newblock 330 (1908), Januar, Nr.~3, S.~377--445

\bibitem[{Monaghan} und {Kocharyan}(1995)]{1995CoPhC..87..225M}
\dinatlabel{{Monaghan} und {Kocharyan} 1995} \textsc{{Monaghan}}, J.~J.~;
  \textsc{{Kocharyan}}, A.:
\newblock {SPH simulation of multi-phase flow}.
\newblock In: \emph{Computer Physics Communications}
\newblock 87 (1995), Mai, Nr.~1-2, S.~225--235

\bibitem[{Moriwaki} und {Nakagawa}(2004)]{2004ApJ...609.1065M}
\dinatlabel{{Moriwaki} und {Nakagawa} 2004} \textsc{{Moriwaki}}, Kazumasa~;
  \textsc{{Nakagawa}}, Yoshitsugu:
\newblock {A Planetesimal Accretion Zone in a Circumbinary Disk}.
\newblock In: \emph{\apj}
\newblock 609 (2004), Juli, Nr.~2, S.~1065--1070

\bibitem[{Morrison} und {Malhotra}(2015)]{2015ApJ...799...41M}
\dinatlabel{{Morrison} und {Malhotra} 2015} \textsc{{Morrison}}, Sarah~;
  \textsc{{Malhotra}}, Renu:
\newblock {Planetary Chaotic Zone Clearing: Destinations and Timescales}.
\newblock In: \emph{\apj}
\newblock 799 (2015), Januar, Nr.~1, S.~41

\bibitem[{Murray} und {Dermott}(1999)]{1999ssd..book.....M}
\dinatlabel{{Murray} und {Dermott} 1999} \textsc{{Murray}}, Carl~D.~;
  \textsc{{Dermott}}, Stanley~F.:
\newblock \emph{{Solar System Dynamics}}.
\newblock 1999

\bibitem[{Paardekooper} u.\,a.(2012){Paardekooper}, {Leinhardt}, {Th{\'e}bault}
  und {Baruteau}]{2012ApJ...754L..16P}
\dinatlabel{{Paardekooper} u.\,a. 2012} \textsc{{Paardekooper}}, Sijme-Jan~;
  \textsc{{Leinhardt}}, Zo{\"e}~M.~; \textsc{{Th{\'e}bault}}, Philippe~;
  \textsc{{Baruteau}}, Cl{\'e}ment:
\newblock {How Not to Build Tatooine: The Difficulty of In Situ Formation of
  Circumbinary Planets Kepler 16b, Kepler 34b, and Kepler 35b}.
\newblock In: \emph{\apjl}
\newblock 754 (2012), Juli, Nr.~1, S.~L16

\bibitem[{Petry} und {CASA Development Team}(2012)]{2012ASPC..461..849P}
\dinatlabel{{Petry} und {CASA Development Team} 2012} \textsc{{Petry}}, Dirk~;
  \textsc{{CASA Development Team}}:
\newblock {Analysing ALMA Data with CASA}.
\newblock In: \textsc{{Ballester}}, P. (Hrsg.)~; \textsc{{Egret}}, D. (Hrsg.)~;
  \textsc{{Lorente}}, N.~P.~F. (Hrsg.): \emph{Astronomical Data Analysis
  Software and Systems XXI} Bd.~461, September 2012, S.~849

\bibitem[{Popova} und {Shevchenko}(2013)]{2013ApJ...769..152P}
\dinatlabel{{Popova} und {Shevchenko} 2013} \textsc{{Popova}}, Elena~A.~;
  \textsc{{Shevchenko}}, Ivan~I.:
\newblock {Kepler-16b: Safe in a Resonance Cell}.
\newblock In: \emph{\apj}
\newblock 769 (2013), Juni, Nr.~2, S.~152

\bibitem[{Press} u.\,a.(1992){Press}, {Teukolsky}, {Vetterling} und
  {Flannery}]{1992nrca.book.....P}
\dinatlabel{{Press} u.\,a. 1992} \textsc{{Press}}, William~H.~;
  \textsc{{Teukolsky}}, Saul~A.~; \textsc{{Vetterling}}, William~T.~;
  \textsc{{Flannery}}, Brian~P.:
\newblock \emph{{Numerical recipes in C. The art of scientific computing}}.
\newblock 1992

\bibitem[{Price}(2012)]{2012JCoPh.231..759P}
\dinatlabel{{Price} 2012} \textsc{{Price}}, Daniel~J.:
\newblock {Smoothed particle hydrodynamics and magnetohydrodynamics}.
\newblock In: \emph{Journal of Computational Physics}
\newblock 231 (2012), Februar, Nr.~3, S.~759--794

\bibitem[{Savin} u.\,a.(2019){Savin}, {Shabanov}, {Telegin} und
  {Baranov}]{2019LG..40..1835}
\dinatlabel{{Savin} u.\,a. 2019} \textsc{{Savin}}, G.I.~; \textsc{{Shabanov}},
  B.M.~; \textsc{{Telegin}}, P.N.~; \textsc{{Baranov}}, A.V.:
\newblock {Joint Supercomputer Center of the Russian Academy of Sciences:
  Present and Future}.
\newblock In: \emph{Lobachevskii Journal of Mathematics}
\newblock 40 (2019), November, S.~1853--1862

\bibitem[{Sefilian} u.\,a.(2023){Sefilian}, {Rafikov} und
  {Wyatt}]{2023ApJ...954..100S}
\dinatlabel{{Sefilian} u.\,a. 2023} \textsc{{Sefilian}}, Antranik~A.~;
  \textsc{{Rafikov}}, Roman~R.~; \textsc{{Wyatt}}, Mark~C.:
\newblock {Formation of Gaps in Self-gravitating Debris Disks by Secular
  Resonance in a Single-planet System. II. Toward a Self-consistent Model}.
\newblock In: \emph{\apj}
\newblock 954 (2023), September, Nr.~1, S.~100

\bibitem[{Shakhovskoj} u.\,a.(2005){Shakhovskoj}, {Grinin} und
  {Rostopchina}]{2005Ap.....48..135S}
\dinatlabel{{Shakhovskoj} u.\,a. 2005} \textsc{{Shakhovskoj}}, D.~N.~;
  \textsc{{Grinin}}, V.~P.~; \textsc{{Rostopchina}}, A.~N.:
\newblock {Analysis of the Historical Light Curve of the UX Ori Star CQ Tau}.
\newblock In: \emph{Astrophysics}
\newblock 48 (2005), April, Nr.~2, S.~135--142

\bibitem[{Shakura} und {Sunyaev}(1973)]{1973A&A....24..337S}
\dinatlabel{{Shakura} und {Sunyaev} 1973} \textsc{{Shakura}}, N.~I.~;
  \textsc{{Sunyaev}}, R.~A.:
\newblock {Black holes in binary systems. Observational appearance.}
\newblock In: \emph{\aap}
\newblock 24 (1973), Januar, S.~337--355

\bibitem[{Springel}(2005)]{2005MNRAS.364.1105S}
\dinatlabel{{Springel} 2005} \textsc{{Springel}}, Volker:
\newblock {The cosmological simulation code GADGET-2}.
\newblock In: \emph{MNRAS}
\newblock 364 (2005), Dec, Nr.~4, S.~1105--1134

\bibitem[{Springel} u.\,a.(2001){Springel}, {Yoshida} und
  {White}]{2001NewA....6...79S}
\dinatlabel{{Springel} u.\,a. 2001} \textsc{{Springel}}, Volker~;
  \textsc{{Yoshida}}, Naoki~; \textsc{{White}}, Simon D.~M.:
\newblock {GADGET: a code for collisionless and gasdynamical cosmological
  simulations}.
\newblock In: \emph{New Astronomy}
\newblock 6 (2001), Apr, Nr.~2, S.~79--117

\bibitem[{Stoer} und {Bulirsch}(1980)]{1980.book.....S}
\dinatlabel{{Stoer} und {Bulirsch} 1980} \textsc{{Stoer}}, J.~;
  \textsc{{Bulirsch}}, R.:
\newblock \emph{{Introduction to Numerical Analysis}}.
\newblock 1980

\bibitem[{Takeuchi} und {Artymowicz}(2001)]{2001ApJ...557..990T}
\dinatlabel{{Takeuchi} und {Artymowicz} 2001} \textsc{{Takeuchi}}, Taku~;
  \textsc{{Artymowicz}}, Pawel:
\newblock {Dust Migration and Morphology in Optically Thin Circumstellar Gas
  Disks}.
\newblock In: \emph{\apj}
\newblock 557 (2001), August, Nr.~2, S.~990--1006

\bibitem[{Tripathi} u.\,a.(2017){Tripathi}, {Andrews}, {Birnstiel} und
  {Wilner}]{2017ApJ...845...44T}
\dinatlabel{{Tripathi} u.\,a. 2017} \textsc{{Tripathi}}, Anjali~;
  \textsc{{Andrews}}, Sean~M.~; \textsc{{Birnstiel}}, Tilman~;
  \textsc{{Wilner}}, David~J.:
\newblock {A millimeter Continuum Size-Luminosity Relationship for
  Protoplanetary Disks}.
\newblock In: \emph{\apj}
\newblock 845 (2017), August, Nr.~1, S.~44

\bibitem[{Ubeira Gabellini} u.\,a.(2019){Ubeira Gabellini}, {Miotello},
  {Facchini}, {Ragusa}, {Lodato}, {Testi}, {Benisty}, {Bruderer}, {T.
  Kurtovic}, {Andrews}, {Carpenter}, {Corder}, {Dipierro}, {Ercolano},
  {Fedele}, {Guidi}, {Henning}, {Isella}, {Kwon}, {Linz}, {McClure}, {Perez},
  {Ricci}, {Rosotti}, {Tazzari} und {Wilner}]{2019MNRAS.486.4638U}
\dinatlabel{{Ubeira Gabellini} u.\,a. 2019} \textsc{{Ubeira Gabellini}},
  M.~G.~; \textsc{{Miotello}}, Anna~; \textsc{{Facchini}}, Stefano~;
  \textsc{{Ragusa}}, Enrico~; \textsc{{Lodato}}, Giuseppe~; \textsc{{Testi}},
  Leonardo~; \textsc{{Benisty}}, Myriam~; \textsc{{Bruderer}}, Simon~;
  \textsc{{T. Kurtovic}}, Nicol{\'a}s~; \textsc{{Andrews}}, Sean~;
  \textsc{{Carpenter}}, John~; \textsc{{Corder}}, Stuartt~A.~;
  \textsc{{Dipierro}}, Giovanni~; \textsc{{Ercolano}}, Barbara~;
  \textsc{{Fedele}}, Davide~; \textsc{{Guidi}}, Greta~; \textsc{{Henning}},
  Thomas~; \textsc{{Isella}}, Andrea~; \textsc{{Kwon}}, Woojin~;
  \textsc{{Linz}}, Hendrik~; \textsc{{McClure}}, Melissa~; \textsc{{Perez}},
  Laura~; \textsc{{Ricci}}, Luca~; \textsc{{Rosotti}}, Giovanni~;
  \textsc{{Tazzari}}, Marco~; \textsc{{Wilner}}, David:
\newblock {A dust and gas cavity in the disc around CQ Tau revealed by ALMA}.
\newblock In: \emph{\mnras}
\newblock 486 (2019), Juli, Nr.~4, S.~4638--4654

\bibitem[{Williams} und {Cieza}(2011)]{2011ARA&A..49...67W}
\dinatlabel{{Williams} und {Cieza} 2011} \textsc{{Williams}}, Jonathan~P.~;
  \textsc{{Cieza}}, Lucas~A.:
\newblock {Protoplanetary Disks and Their Evolution}.
\newblock In: \emph{\araa}
\newblock 49 (2011), September, Nr.~1, S.~67--117

\bibitem[{W{\"o}lfer} u.\,a.(2021){W{\"o}lfer}, {Facchini}, {Kurtovic},
  {Teague}, {van Dishoeck}, {Benisty}, {Ercolano}, {Lodato}, {Miotello},
  {Rosotti}, {Testi} und {Ubeira Gabellini}]{2021A&A...648A..19W}
\dinatlabel{{W{\"o}lfer} u.\,a. 2021} \textsc{{W{\"o}lfer}}, L.~;
  \textsc{{Facchini}}, S.~; \textsc{{Kurtovic}}, N.~T.~; \textsc{{Teague}},
  R.~; \textsc{{van Dishoeck}}, E.~F.~; \textsc{{Benisty}}, M.~;
  \textsc{{Ercolano}}, B.~; \textsc{{Lodato}}, G.~; \textsc{{Miotello}}, A.~;
  \textsc{{Rosotti}}, G.~; \textsc{{Testi}}, L.~; \textsc{{Ubeira Gabellini}},
  M.~G.:
\newblock {A highly non-Keplerian protoplanetary disc. Spiral structure in the
  gas disc of CQ Tau}.
\newblock In: \emph{\aap}
\newblock 648 (2021), April, S.~A19

\end{thebibliography}
\bibliographystyle{dinat} 
\end{document}